# Ionic conductivity of deep eutectic solvents: The role of orientational dynamics and glassy freezing

Daniel Reuter, Catharina Binder, Peter Lunkenheimer* and Alois Loidl

We have performed a thorough examination of the reorientational relaxation dynamics and the ionic charge transport of three typical deep eutectic solvents, ethaline, glyceline and reline by broadband dielectric spectroscopy. Our experiments cover a broad temperature range from the low-viscosity liquid down to the deeply supercooled state, allowing to investigate the significant influence of glassy freezing on the ionic charge transport in these systems. In addition, we provide evidence for a close coupling of the ionic conductivity in these materials to reorientational dipolar motions which should be considered when searching for deep eutectic solvents optimized for electrochemical applications.

## 1 Introduction

In recent years, the so-called deep eutectic solvents (DESs) have attracted tremendous interest, being considered as a superior alternative to ionic liquids (ILs) concerning ease of preparation, low cost, sustainability, biocompatibility and low toxicity.[1,2,3,4,5,6] ILs essentially can be described as salts that are liquid close to room temperature. In contrast to ILs, DESs are multicomponent systems and, in their most common form, are composed of a salt (often a quaternary ammonium salt) and a molecular hydrogen-bond donor as glycerol or urea. Often these constituents are also found in nature, for which the term "natural DES" was coined, emphasizing their environmental friendliness and renewability.[3,5,7] The mixing of the two components generates the typical melting-point reduction at the eutectic composition, which makes the DES liquid at room temperature. This in principle can transform salts into an IL-like state that otherwise are solid at room temperature and thus unsuited as solutes or electrolytes. Just as ILs, DESs are promising new materials for a large number of applications and can be used, e.g., as "green solvents" for synthesis and material preparation, as extraction agents, or for biocatalysis.[2,3,4,6] As they are ionic solvents, it seems natural to also consider DESs for electrochemical applications, using them as electrolytes in batteries and solar cells, for the electrodeposition of metals, or for electropolishing of metal surfaces.[2,4,8,9,10,11]

For ionically conducting materials, dielectric spectroscopy is an ideal tool to learn more about the ionic charge-transport mechanism and possible additional dynamic processes. This experimental method not only provides information on the ion motion, via the detection of the dc and ac conductivity, but also on possible reorientational degrees of freedom of dipolar molecules and ions.[12,13] The latter dynamics, which usually leads to characteristic relaxational response in the spectra, termed $\alpha$ relaxation, often is coupled with the viscosity to varying extents[13,14,15] and even can considerably influence the ion mobility of certain materials.[16,17,18]

When measured in a sufficiently broad temperature range, the dc conductivity of DESs usually exhibits pronounced non-Arrhenius behaviour,[10,11,19] typical for glassforming liquids. Dielectric spectroscopy is also ideally suited to investigate the glassy freezing in such materials by monitoring the non-canonical continuous slowing down of their internal degrees of freedom in a broad temperature and dynamic range.[12,13,20] For ILs, it is well established that their glass transition plays an important role for the value of the ionic dc conductivity at room temperature, which is influenced by both the glass-transition temperature[21] and the so-called fragility parameter,[22] the latter providing a measure of the deviation from Arrhenius behaviour.[23] However, not much is known about the role of the glass transition in DESs.

While numerous and extensive investigations of ILs using dielectric spectroscopy have been reported in literature (e.g., Refs. 18,22,24,25,26,27,28,29,30,31,32,33,34), only few such studies were performed for DESs.[19,35,36,37] Moreover, most of these investigations either were restricted to frequencies around several GHz,[35,37] not allowing to trace the glassy freezing when approaching the glass state, or concentrated on the translational ion dynamics by primarily analysing the dielectric data in terms of the modulus formalism.[19] For ILs, our group has recently pointed out the importance of *reorientational* motions for this class of materials[18], which were previously only rarely treated[25,28,38,39] We speculated that these reorientations may even be of comparable significance for the ionic charge transport in ILs as in ionically conducting plastic crystals. In the latter, the centres of mass of the molecules are located on a crystalline lattice, but they are dynamically disordered with respect to the orientational degrees of freedom.[40] This may open paths for ionic charge transport via a revolving-door or paddle-wheel-like mechanism.[16,17,41,42] For DESs, however, the role of reorientational motions still remains to be clarified.

In the present work, we investigate the dielectric permittivity and conductivity of two typical DESs, ethaline and reline, and further analyse experimental data on glyceline that were published previously.[36] These are mixtures of the salt choline chloride (ChCl) with ethylene glycol, urea or glycerol, respectively. Choline and the latter three, acting as hydrogen donors, represent asymmetric molecules and, thus, their reorientational motions should be detectable by dielectric spectroscopy. Indeed, for pure urea[37], ethylene glycol[43,44,45] and glycerol[20,46] the corresponding relaxational dielectric response is well known and the polar nature of choline-based DESs was pointed out, based on solvatochromic optical measurements.[47] In the case of glycerol, which is a very good and often investigated glass former, the relaxation dynamics can be traced over an extremely broad frequency/time range, covering temperatures from the low-viscosity liquid down into the solid-glass state.[20,48] Here we concentrate on the investigation of the so far only incompletely investigated *reorientational* dynamics in DESs and its glassy freezing. We especially address the influence of the glass transition on the ionic conductivity and the coupling of reorientational molecular and translational ionic motion in these materials. For this purpose, we have performed dielectric spectroscopy in a broad frequency range between 0.1 Hz and about 3 GHz, covering a wide temperature range between the liquid and the deeply supercooled state close to the glass transition.

*Experimental Physics V, Center for Electronic Correlations and Magnetism, University of Augsburg, 86135 Augsburg, Germany.*
*E-mail: peter.lunkenheimer@physik.uni-augsburg.de*
† Electronic supplementary information (ESI) available: Fig. S1.

## 2 Experimental details

The chemicals urea (purity > 99 %), choline chloride (purity > 98 %) and ethylene glycol (purity > 99 %) were purchased from Alfa Aesar and used as received. The DESs ethaline (ChCl + ethylene glycol, 1:2 molar ratio) and reline (ChCl + urea, 1:2 molar ratio) were prepared by mixing appropriate amounts of the two components inside a glass tube using a magnetic stirrer for 24 hours at 350 K. In both cases, the result was a colourless liquid without residual crystalline particles. Before the dielectric and calorimetric measurements, the water content of both samples was tested by coulometric Karl-Fischer-Titration and determined to be significantly smaller than 1 wt%. At 350 K, the reline mixture was found to exhibit a dc conductivity of about 8 mS/cm, which is of similar order as the values reported by Abbott *et al.*[1] and Du *et al.*[49]

Differential scanning calorimetry (DSC) measurements were performed using a DSC 8500 from Perkin Elmer and a scanning rate of 10 K/min during heating and cooling. From the DSC data, glass-transition temperatures were determined as the onset of the step-wise increase at the glass transition during heating (Fig. S1 in ESI†).

To obtain broadband dielectric data, covering a frequency range of 0.1 Hz to about 3 GHz, a frequency-response analyser (Novocontrol Alpha-A analyzer) and an impedance analyser (Keysight Technologies E4991B) with coaxial reflectometric setup[50] were used. Stainless steel was chosen as capacitor material for all measurements to account for the relatively high chemical reactivity of the DES.[4] For measurements at frequencies up to 1 MHz, the use of parallel-plate capacitors with rather large plate distances, $d > 1$ mm, proved advantageous: Large $d$ essentially leads to a shift of the extrinsic, surface-related contributions to lower frequencies[51] and, thus, results in a better separation from the intrinsic contributions in the dielectric spectra. Temperature regulation was carried out via a N$_2$-gas flow cryostat (Novocontrol Quatro).

## 3 Results and Discussion

Figure 1 shows the obtained spectra of the real and imaginary part of the permittivity (the dielectric constant $\varepsilon'$ (a) and the dielectric loss $\varepsilon''$ (b), respectively) and of the real part of the conductivity $\sigma'$ (c) as revealed by ethaline for a variety of temperatures. The investigated temperatures cover the range from the low-viscosity liquid above room temperature down to the deeply supercooled liquid state, about 10 K above the glass-transition temperature $T_g \approx 155$ K as determined by DSC (Fig. S1 in ESI†). The strong increase of $\varepsilon'(\nu)$, approaching extremely large values, observed in Fig. 1(a) for low frequencies and the higher temperatures, clearly is due to non-intrinsic, electrode-related effects (sometimes termed "blocking electrodes"). They are typical for ionically conducting materials (e.g., ILs[18]) and arise when the mobile ions arrive at the electrodes at low frequencies, forming thin, poorly conducting space-charge regions, acting as huge capacitors.[52] Huge values of $\varepsilon'$ at low frequencies (however, ascribed to molecular aggregation) were also detected in an early dielectric investigation of a DES in Ref. 53. It should be noted that these blocking layers observed in dielectric spectroscopy are closely related to the so-called double-layer capacitance, relevant for certain technical applications of ionic conductors (e.g., supercapacitors), which is often investigated for their electrochemical characterization (see Refs. 4 and 54 for examples in DESs).

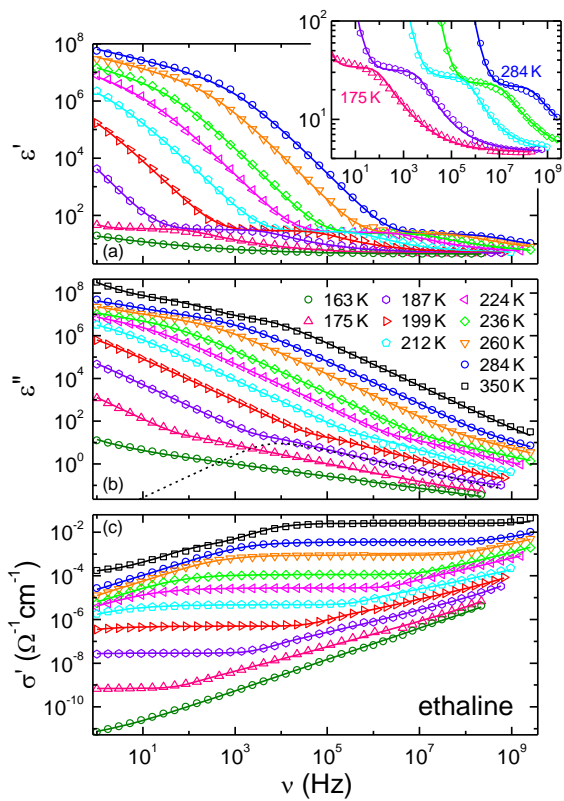

Fig. 1 Spectra of the dielectric constant $\varepsilon'$ (a), the dielectric loss $\varepsilon''$ (b) and the real part of the conductivity $\sigma'$ (c) as measured at various temperatures for ethaline. The inset in (a) shows a zoomed view of the relaxation steps. The solid lines in (a) and (b) are fits assuming up to two distributed RC circuits to model the blocking electrodes, one relaxational process and a contribution from dc conductivity (see text for details). $\varepsilon'(\nu)$ and $\varepsilon''(\nu)$ were simultaneously fitted. The dashed line in (b) indicates the contribution of the $\alpha$ relaxation for 187 K. The lines in (c) were calculated from the fits of $\varepsilon''$ via $\sigma' = \varepsilon'' \varepsilon_0 \omega$.

At frequencies beyond the regime of this non-intrinsic contribution, the $\varepsilon'$ spectra exhibit a well-pronounced step-like decrease with increasing frequency, strongly shifting to lower frequencies with decreasing temperature. It is more clearly revealed in the inset of Fig. 1(a) showing a zoomed view for selected temperatures. This is the typical signature of a relaxational process[12,13,20] and we ascribe it to the reorientational motion of the dipolar molecules in this DES. Similar relaxation processes were previously only detected in few other DESs, moreover in rather restricted temperature and/or frequency ranges only.[35,37,53] In principle, the $\varepsilon'$ spectra shown in the inset of Fig. 1 closely resemble those of conventional dipolar liquids under supercooling,[12,20,48,55] which could be expected when considering that 2/3 of the samples indeed consist of such a dipolar liquid. The only difference to $\varepsilon'$ spectra of conventional liquids is the additional contribution from electrode polarization at low frequencies, arising from the mobility of the added ions. Therefore, it seems natural to analyse these data in terms of the permittivity instead of the modulus representation,[56] often employed for ionic conductors. Since the pioneering works by Debye, it is clear that in this way significant information about reorientational dipolar motions can be obtained.

While two dipolar molecules (choline and ethylene glycol) exist in this DES, obviously only a single dipolar $\alpha$-relaxation process is observed in Fig. 1(a). It is well known that liquids comprising two dipolar components often exhibit a single $\alpha$ relaxation[57,58] Obviously, such binary dipole mixtures represent a new material with its own dipolar dynamics, no longer reflecting the reorientational motion of the two individual dipole species.

In dielectric loss spectra, a dipolar relaxational process should lead to a peak, located at a frequency close to the point of inflection of the corresponding step in $\varepsilon'(\nu)$. However, in Fig. 1(b) only the right flanks of the expected peaks and a slight indication of a shoulder are visible (cf. dashed line in Fig. 1(b), indicating the $\alpha$-relaxation peak for 187 K). Their low-frequency flanks are obscured by a strong increase with slope -1 in this double-logarithmic plot, i.e., $\varepsilon''$ varies like $\nu^1$ in this region. When considering the general relation $\varepsilon'' \propto \sigma'/\nu$, it is obvious that this feature reflects a constant conductivity contribution. The latter indeed directly shows up as a plateau in the $\sigma'$ spectra shown in Fig. 1(c) and signifies the ionic dc charge-transport in this DES. The continuous rise of $\sigma'(\nu)$ following this plateau at higher frequencies and completely dominating the spectrum at the lowest temperature is due to the relaxational response as also seen in the loss spectra discussed above. The reduction of $\sigma'(\nu)$ with decreasing frequency, showing up at high temperatures in Fig. 1(c) (e.g., at frequencies below about $10^4$ Hz for the 350 K curve), arises from the formation of blocking electrodes.[52]

It should be noted that some models on translational charge transport via ion hopping predict a relaxation-like step in $\varepsilon'(\nu)$, which is ascribed to local ion motion. The most prominent one is the random free-energy barrier hopping model (RBM),[59] which was applied to various ILs[31,60,61] and also to two DES systems.[19] In Fig. 2, as an example we show $\varepsilon'(\nu)$ and $\sigma'(\nu)$ for ethaline at 199 K, fitted with the RBM and an extra contribution, accounting for the electrode-polarization as explained below. Obviously, the experimental data cannot be described in this way. When accepting the notion that the observed step in $\varepsilon'(\nu)$ is due to reorientational molecular motions, it is clear that the RBM, exclusively considering translational ion motions, cannot be expected to fit such data. As mentioned above, the increase of $\sigma'(\nu)$ at high frequencies observed in these DESs also is completely due to molecular relaxation while in the RBM it is ascribed to ionic hopping conductivity. It is thus clear that the RBM should not be applied to DESs with strong dipolar relaxation. The only visible contributions from ion hopping to these spectra are the dc plateau in $\sigma'(\nu)$ and the electrode effects seen in both quantities at low frequencies. Any possible ac conductivity contributions from hopping, if present at all, must be hidden under the dominating relaxation features.

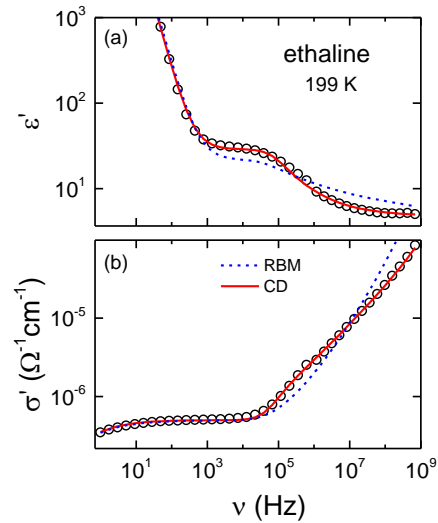

Fig. 2 Frequency dependence of $\varepsilon'$ (a) and $\sigma'$ (b) of ethaline, measured at 199 K. The solid lines are fits using a CD function for the $\alpha$-relaxation as already shown in Fig. 1. The dashed line represents a fit using the RBM.[59] In both cases, the electrode polarization was taken into account by a distributed RC circuit, connected in series to the bulk.[52] The fits were simultaneously performed for $\varepsilon'$ and $\sigma'$.

The spectra of reline ($T_g$ from DSC $\approx$ 205 K; Fig. S1 in ESI†) presented in Fig. 3, in most respects exhibit qualitatively similar behaviour as those of ethaline. Especially, there is clear evidence of a dipolar relaxation process as revealed by the step in $\varepsilon'(\nu)$ shown in Fig. 3(a) in a zoomed representation. A similar statement can also be made for the spectra on glyceline (an 1:2 mixture of ChCl and glycerol; $T_g \approx$ 175 K) published in the Supporting Information of Ref. 36. However, in reline the signature of the relaxation peaks in $\varepsilon''(\nu)$ seems less pronounced than in the two other systems, which indicates a higher relative amplitude of the conductivity contribution (cf. dashed line in Fig. 3(b)). Moreover, in reline a weak loss peak is revealed at the lowest temperature (at about $10^5$ Hz for 212 K in Fig. 3(b)). It clearly is not related to the $\alpha$ relaxation, which has shifted out of the frequency window for this temperature (cf. Fig. 3(a)). Such secondary relaxation processes that are faster and weaker than the $\alpha$ relaxation are common phenomena of dipolar glassforming liquids[62,63,64] and were also found for ILs.[18,26] Interestingly, in the dielectric modulus spectra of two DESs reported in Ref. 19, secondary processes were also detected. A detailed treatment of this spectral feature, however, is out of the scope of the present work. Again, we found that the spectra of Fig. 3 cannot be satisfactorily fitted by the RBM model, in accord with the reorientational origin of the detected relaxational response.

To extract the intrinsic relaxation parameters, we performed simultaneous fits of the $\varepsilon'$ and $\varepsilon''$ data of Figs. 1 and 3. To account for the non-intrinsic electrode-polarization effects, affecting the spectra at low frequencies and high temperatures, we employed an equivalent circuit with up to two distributed RC circuits, connected in series to the sample.[52] Such an empirical approach was previously demonstrated to enable a proper description of electrode effects in various types of ionic conductors, including ILs and plastic crystals.[17,18] Only at the highest temperatures shown in Figs. 1 and 3, two instead of a single RC circuit had to be used. This accounts for the fact that there the mentioned blocking-induced conductivity reduction at low frequencies exhibits two successive steps (at lower temperatures, the low-frequency step has shifted out of the frequency window). A similar two-step behaviour at high temperatures was also detected in the conductivity spectra of two DESs published in Ref. 19. Possible microscopic origins of such behaviour were discussed, e.g., in Ref. 52.

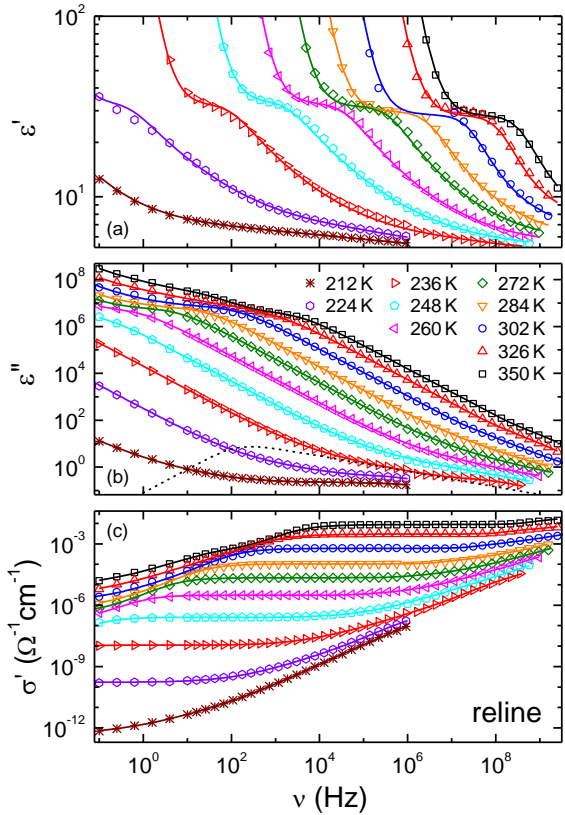

achieved in this way (solid lines in Figs. 1 and 3). It should be noted that, depending on temperature, only part of the different contributions had to be employed to fit the different spectra, restricting the overall number of parameters to reasonable values. For example, the electrode effects play no role at low temperatures (there they would occur at much lower frequencies than covered by the experiments) while the secondary relaxation in reline is not needed at high temperatures, were it has shifted beyond the investigated maximum frequency. Notably, in the high-frequency investigations of several acetamide-based DESs in the 0.2 - 50 GHz range, up to four Debye relaxation functions had to be used to fit the dielectric spectra around room temperature.[35,37] In contrast, for the present data up to about 1 GHz, we found that a single broadened relaxation function is sufficient to fit the spectra, at least at high temperatures, where the secondary relaxation plays no role. It is clear that the very fast processes detected in Refs. 35 and 37, if present at all in our systems, would mostly be outside of our frequency window.

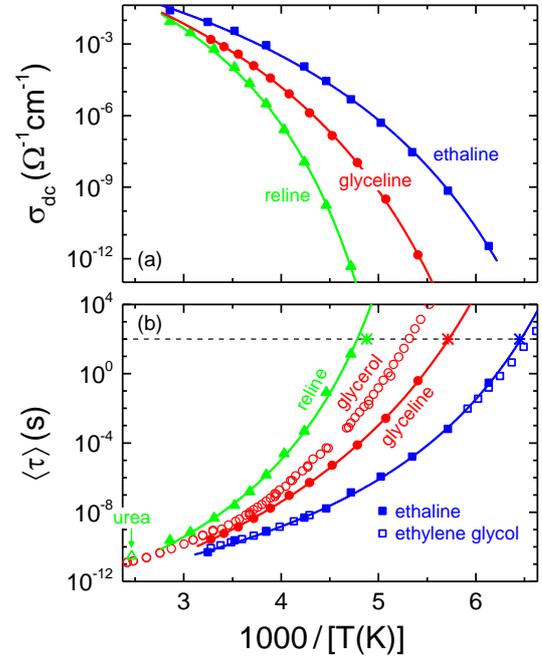

Fig. 3 Spectra of the dielectric constant $\varepsilon'$ (a), the dielectric loss $\varepsilon''$ (b) and the real part of the conductivity $\sigma'$ (c) as measured at various temperatures for reline. The lines in (a) and (b) are fits assuming up to two distributed RC circuits to model the blocking electrodes, two relaxational processes and a contribution from dc conductivity (see text for details). $\varepsilon'(\nu)$ and $\varepsilon''(\nu)$ were simultaneously fitted. The dashed line in (b) indicates the contribution of the $\alpha$ relaxation for 236 K. The lines in (c) were calculated from the fits of $\varepsilon''$ via $\sigma' = \varepsilon'' \varepsilon_0 \omega$.

For both systems, we employed the Cole-Davidson (CD) function to fit the intrinsic $\alpha$-relaxation process.[46] The secondary relaxation, clearly revealed in the reline loss spectra at low temperatures, was described by a Cole-Cole (CC) function.[65] Both are commonly applied empirical functions for the $\alpha$ and secondary relaxations in supercooled dipolar liquids, respectively[13,20,64] They account for the broadened spectral shape of the relaxation features, usually termed "non-exponentiality" and ascribed to dynamical heterogeneity.[66,67] For both materials, the mentioned contribution of the dc conductivity $\sigma_{dc}$ to the loss was taken into account by a term $\varepsilon''_{dc} = \sigma_{dc} / (\varepsilon_0 \omega)$ (with $\varepsilon_0$ the permittivity of free space and $\omega = 2\pi\nu$). Overall, good fits of the experimental data could be

Fig. 4 (a) Arrhenius representation of the temperature-dependent dc conductivity of the three DESs treated in this work as resulting from the fits of the dielectric spectra. The solid lines are fits with the VFT formula, eqn. (1). (b) Arrhenius representation of the temperature-dependent reorientational relaxation times of the investigated DESs (closed symbols) and the corresponding hydrogen-bond donors (open symbols). The DES data were determined from the fits of the dielectric spectra. Data on glycerol were taken from Ref. 82 and the single data point for molten urea from Ref. 37; data on ethylene glycol were determined in the present work. The lines in (b) are fits with eqn. (2). The horizontal dashed line indicates $\tau(T_g) = 100$ s. The stars are based on the estimates of $T_g$ from the DSC measurements (Fig. S1 in ESI†).

Figure 4(a) shows an Arrhenius representation of the temperature-dependent dc conductivity as resulting from the fits of the dielectric spectra for ethaline (Fig. 1), reline (Fig. 3) and glyceline.[36] We find significant differences in the conductivity of these three DESs, least pronounced close to room temperature but reaching many decades at low temperatures, with ethaline being the best conductor. The covered broad temperature and conductivity range, which significantly exceeds that of previous investigations[1,49,68,69] clearly reveals strong deviations of $\sigma_{dc}(T)$ from Arrhenius behaviour. Such deviations from simple thermally activated ionic charge transport are common findings for glass-forming ionic conductors[17,18,22,70] and usually can be well described by a modification of the empirical Vogel-Fulcher-Tammann (VFT) law:[71,72,73]

$$\sigma_{dc} = \sigma_0 \exp\left[\frac{-D_\sigma T_{\mathrm{VF}\sigma}}{T - T_{\mathrm{VF}\sigma}}\right] \quad (1)$$

Here $\sigma_0$ is a pre-exponential factor, $D_\sigma$ is the so-called strength parameter[23] and $T_{\mathrm{VF}\sigma}$ is the Vogel-Fulcher temperature, where the conductivity would become zero. The lines in Fig. 4(a) are fits with eqn. (1) (see Table 1 for the parameters), leading to a reasonable description of the experimental data. VFT behaviour of $\sigma_{dc}$ was previously also reported for other DESs[10,11,19,68] but, to our knowledge, only in a single case[19] it was traced over a similarly broad temperature and conductivity range as in the present work.

Table 1 Parameters obtained from the VFT fits of $\sigma_{dc}(T)$ and $\langle\tau\rangle(T)$ shown in Fig. 4 and the fragility parameter calculated from $D_\tau$.

|  | $T_{\mathrm{VF}\sigma}$(K) | $D_\sigma$ | $\sigma_0$ ($\Omega^{-1}$cm$^{-1}$) | $T_{\mathrm{VF}\tau}$(K) | $D_\tau$ | $\tau_0$ (s) | $m$ |
|---|---|---|---|---|---|---|---|
| ethaline | 111 | 13.7 | 81 | 113 | 13.2 | $2.8\times10^{-14}$ | 60 |
| glyceline | 123 | 16.0 | 19 | 113 | 21.3 | $8.5\times10^{-16}$ | 47 |
| reline | 159 | 11.0 | 112 | 152 | 14.3 | $2.3\times10^{15}$ | 57 |

The second important quantity resulting from the fits of the dielectric spectra is the reorientational $\alpha$-relaxation time. The temperature dependence of the average relaxation time $\langle\tau\rangle$, calculated from the obtained parameters of the CD function, is shown in Fig. 4(b) for the three investigated DESs (closed symbols). Again, marked non-Arrhenius behaviour is revealed which can be well fitted by the VFT law (see Table 1 for parameters):

$$\langle\tau\rangle = \tau_0 \exp\left[\frac{D_\tau T_{\mathrm{VF}\tau}}{T - T_{\mathrm{VF}\tau}}\right] \quad (2)$$

Such VFT behaviour of $\tau(T)$ is a characteristic property of glassforming dipolar liquids[13,20,48,74,75] and mirrors the non-canonical freezing of molecular motions when approaching the glass transition. While the underlying microscopic mechanism is far from being finally clarified, in recent years evidence has been growing[76,77,78] that old concepts, assuming an increasing length scale of molecular cooperativity due to an underlying phase transition,[79] are the correct explanation. From $D_\tau$ the fragility parameter can be calculated (Table 1).[80] It provides a measure of the deviations of $\tau(T)$ from Arrhenius temperature dependence with $m = 16$ implying no deviations (so-called "strong" glass formers) and $m = 170$ extremely "fragile" behaviour.[81] The found values between 47 and 60 characterize the present DESs as intermediate within this classification system, similar to glycerol ($m = 53$)[80] forming the main compound of glyceline. From temperature-dependent relaxation-time data, the glass-transition temperature can be estimated via the relation $\tau(T_g) \approx 100$ s. An extrapolation of the VFT fit curves in Fig. 4(b) leads to values of 155, 175 and 209 K for ethaline, glyceline and reline, respectively. These reasonably agree with the glass temperatures determined by DSC (155, 175, and 205 K, respectively), which are indicated by the stars in Fig. 4(b).

For comparison, Fig. 4(b) also shows the $\alpha$-relaxation times of the three hydrogen-bond donors representing the main components of the investigated DESs (open symbols).[37,82] As already pointed out in Ref. 36, at low temperatures glyceline exhibits significantly faster relaxational dynamics than pure glycerol. This agrees with the finding that adding ChCl to glycerol decreases the viscosity of the system.[83] Obviously, the breaking and reforming of the hydrogen-bond network that is necessary for reorientational processes (and viscous flow) in glycerol is strongly modified by the addition of ChCl. One may ascribe this finding to the partial breaking of the three-dimensional hydrogen network in the liquid and supercooled states of glycerol by adding ions. However, one should note that, e.g., the addition of LiCl to glycerol leads to opposite behaviour, i.e., a slowing down of the reorientational dynamics.[84] In the latter system, the glycerol molecules represent the only dipolar species while the choline ions in glyceline also have a dipolar moment. Hence, the change in relaxation time strongly depends on the added ion species, just as shown for the viscosity of other glycerol/ion mixtures.[85] In any case, the approach of very similar $\tau$ values for glycerol and glyceline, observed at high temperatures in Fig. 4(b), strongly confirms the orientational origin of the relaxation process in the latter, for which reorientations of the glycerol molecules certainly play an important role.

Interestingly, in ethaline the situation is different than in glyceline: Comparing the open and closed squares in Fig. 4(b) reveals that its relaxation times closely agree with those of its main constituent, ethylene glycol, in the whole temperature range. One may speculate that this different behaviour is somehow related to the existence of only two OH groups in the ethylene glycol molecule, instead of three in glycerol, which may lead to a less three-dimensional molecular network, less affected by partial breaking. It is clear that dielectric measurements are not able to finally solve these open questions, which may be tackled by future nuclear magnetic resonance or structural investigations.

Urea, the main component of reline, unfortunately has not been investigated in its supercooled state until now, which may be difficult to access, requiring unrealistically fast quenching. The single data point of molten urea shown in Fig. 4(b)[37] well agrees with the extrapolated $\tau(T)$ of reline. As the relaxation times of glycerol and glyceline also approach each other at high temperatures, for urea and reline no statement on a possible deviation of their relaxation times at low temperatures can be made.

As evidenced by Fig. 4, both the ionic charge transport and the reorientational motion in the investigated DESs show VFT behaviour and, thus, both are obviously governed by the glassy freezing of these materials. Therefore, it seems likely that these dynamics are closely coupled. The VFT parameters $D$ and $T_{\mathrm{VF}}$ for $\sigma_{dc}(T)$ and $\tau(T)$ indeed are of comparable order (Table 1). A more direct check of the coupling of conductivity and reorientation is provided by Fig. 5: It presents an Arrhenius plot of the temperature-dependent dc resistivity $\rho_{dc} = 1/\sigma_{dc}$ (circles) and of $\langle\tau\rangle$ (crosses) in a common frame (we plot $\rho_{dc}$ instead of $\sigma_{dc}$ because it has a negative temperature gradient just as $\tau$, enabling a direct comparison). Here the two ordinates were adjusted to cover the same number of decades. By a proper choice of the starting values of the y-axes, a nearly perfect match of the resistivity and relaxation-time curves can be achieved, at least for ethaline and glyceline. This implies a direct proportionality of both quantities. For reline, minor deviations show up at low temperatures but the overall temperature dependence of both quantities still is quite similar, pointing to a close coupling.

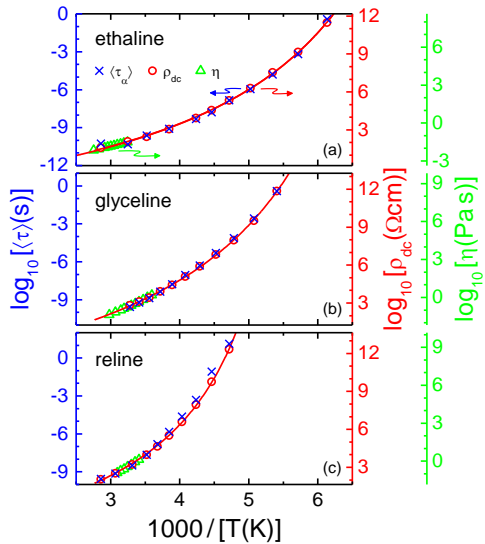

Fig. 5 Comparison of the temperature-dependent translational and reorientational dynamics of the three investigated DESs. The average reorientational $\alpha$-relaxation times from dielectric spectroscopy (crosses; left scale) and the dc resistivity $\rho_{dc}$ (circles; first right scale) are shown in Arrhenius representation. If available, in addition viscosity data[68,86,87] (triangles; second right scale) are included. The different ordinates cover the same number of decades. Their starting values were chosen to achieve a good match of the different quantities. The lines are VFT fits of $\rho_{dc}$.

The question arises whether this coupling is of direct nature, e.g., via a revolving-door mechanism as considered for plastic crystals[16,17,41,42] or even for ILs,[18] or whether it is mediated by the viscosity $\eta$, to which both dynamics may directly couple. A comparison with viscosity data of these materials could help to solve this question. However, unfortunately such data only exist in a rather restricted temperature range around room temperature as shown by the triangles in Fig. 5.[68,86,87] Here, just as described above for $\tau$ and $\rho_{dc}$, the $\eta$ data were adjusted to match the two other data sets. This works nearly perfectly, implying $\eta \propto \tau \propto \rho_{dc}$ at least at high temperatures. However, presently no statement can be made about a possible decoupling of $\eta$ from the two other quantities at low temperatures as reported for some ILs in Ref. 18.

To further examine the relation of the ionic charge transport and molecular reorientations, Fig. 6(a) shows $\rho_{dc}$ vs. $\langle\tau\rangle$ for the three DESs in double-logarithmic representation. As indicated by the solid line with slope one, both for ethaline and glyceline indeed proportionality of these quantities is found, a behaviour sometimes referred to as Debye-Stokes-Einstein relation.[88,89] Moreover, the $\rho_{dc}(\tau)$ curves of both DESs agree within experimental resolution. This implies that a certain reorientation rate of the dipolar molecules induces a similar resistivity value, independent of the type of the reorienting hydrogen-bond donor molecule (glycerol or ethylene glycol). In marked contrast, for reline lower resistivity than in the other systems is found for the same relaxation times, the difference reaching more than a factor of 10 at high $\tau$ (i.e., at low temperatures). Moreover, $\rho_{dc}$ and $\tau$ exhibit a small but significant deviation from proportionality as indicated by the dashed line in Fig. 6 with slope 0.93, implying $\rho_{dc} \propto \langle\tau\rangle^{0.93}$. (This reminds of the fractional Stokes-Einstein or fractional Debye-Stokes-Einstein behaviour found in various other classes of glassforming systems, however, usually related to quantities like viscosity and various diffusion coefficients.[89,90,91,92]) It seems reasonable to ascribe the significantly lower resistivity of reline for given relaxation times to this decoupling of ionic translational and reorientational mobility. Within the revolving-door mechanism, this would imply that in reline the mobile ions (probably Cl⁻) find additional paths that allow bypassing the "doors" formed by the urea molecules and possibly also by the less mobile dipolar choline ions. The lower resistivity, i.e., higher conductivity in relation to its reorientational dynamics, found in reline, also explains why, in its loss spectra (Fig. 3(b)), the relaxation peaks are much more strongly obscured by the dc conductivity than for ethaline (Fig. 1(b)). Finally, it should be mentioned that various microscopic explanations of translation-rotation decoupling found in other types of glassforming liquids were proposed (e.g., Refs. 90,91,92,93)

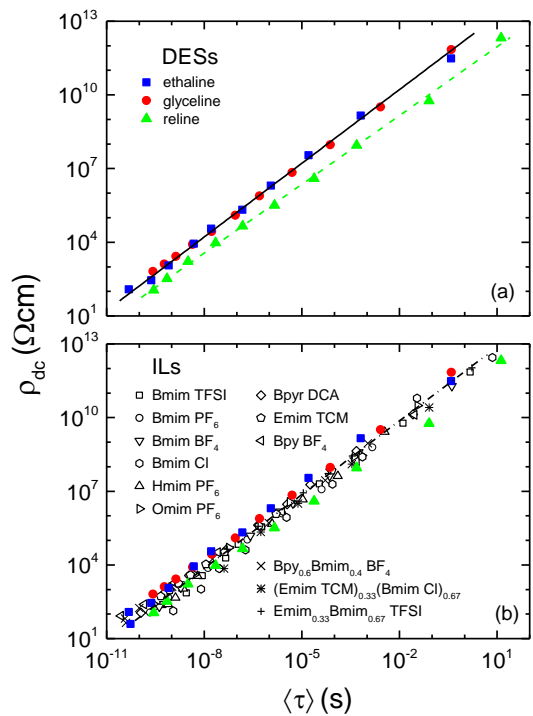

Fig. 6 (a) Dependence of the ionic dc resistivities of the three investigated DESs on their reorientational relaxation times. The solid line with slope one indicates linear behaviour, $\rho_{dc} \propto \langle\tau\rangle$ for ethaline and glyceline. The dashed line has slope 0.93, implying $\rho_{dc} \propto \langle\tau\rangle^{0.93}$ for reline. (b) The same plot as (a), also including data on various ILs from Ref. 18. The line indicates linear behaviour (slope 1).

For comparison, Fig. 6(b) shows the same $\rho_{dc}$ vs. $\langle\tau\rangle$ plot as in frame (a), also including data on various ILs from Ref. 18. All these systems exhibit rather similar behaviour as the present three DESs and their $\rho_{dc}$ for given $\tau$ varies within about a single decade only, with reline still the "best" concerning its relative dc resistivity at low temperatures. It is rather astonishing that such different materials, partly comprising ions as sole dipolar species (the ILs) and partly with additional dipolar molecules (the DESs), exhibit such similar coupling of ionic translation and reorientational motions.

Notably, while for the same *relaxation time* reline exhibits the highest conductivity of the three investigated DESs (Fig. 6(a)), for the same *temperature*, in fact it has the lowest conductivity as becomes obvious from Fig. 4(a). However, without the found decoupling phenomenon, its conductivity would be even lower. This demonstrates the relevance of the translation-rotation coupling investigated in the present work for the absolute value of the technically relevant dc conductivity around room temperature. Obviously, the second important factor determining the conductivity of these DESs is their glass-transition temperature which is the highest for reline. As become immediately clear from an inspection of Fig. 4(b), the absolute value of $T_g$ strongly affects the relaxation time at room temperature, which varies by almost a factor of 100 for the three systems. Due to the close coupling of $\sigma_{dc}$ and $\tau$, the dc conductivity at room temperature also is governed by $T_g$, very similar to the findings in ILs.[21] However, there is also some limited influence of the fragility, determining the bending of the curves in Fig. 4(a).[22]

## 4 Conclusions

In the present work, we have provided a thorough characterization of two typical DESs, ethaline and reline, using dielectric spectroscopy, covering the complete range from the low-viscosity liquid down to the deeply supercooled state close to the glass transition. We especially have concentrated on their only scarcely investigated reorientational molecular dynamics,[35,36,37] which must be present in these systems when considering the dipolar nature of their main constituents. By analysing dielectric permittivity spectra in a broad frequency range, including those on the DES glyceline published previously, we find clear evidence for relaxational processes that arise from such reorientational motions. These processes exhibit clear signatures of glassy freezing as non-Arrhenius and non-exponential relaxation.[74,75] At high temperatures, the relaxational dynamics of these DESs seems to be similar as for the pure dipolar liquids representing their main constituents. However, at low temperatures glyceline relaxes much faster than glycerol which is not the case for ethaline and ethylene glycol. The microscopic origin of this puzzling difference still needs to be clarified.

Most importantly, we find a close coupling of the ionic translational motions, quantified by the dc conductivity, to the reorientational motions in these DESs. Further work, especially investigating the viscosity in larger temperature regions, are necessary to clarify the question whether this coupling arises via a revolving-door mechanism, as considered for other classes of ionic conductors with reorientational degrees of freedom,[16,17,18] or whether it arises from an equal coupling of both dynamics to the viscosity. Obviously, the dc conductivity at room temperature is governed by both, this translational-rotational coupling and the glassy freezing of these systems as characterized by their glass temperature and fragility. In the search for DESs for electrochemical applications, the optimization of the dc conductivity is one of the most important tasks. The results of the present work indicate that this can be achieved in several ways: by lowering the glass temperature, by increasing the fragility and by reducing the translational-rotational coupling.

## Conflicts of interest

There are no conflicts to declare.

## Acknowledgements

This work was supported by the Deutsche Forschungsgemeinschaft (Grant No. LU 656/3-1).

## Notes and references

# Electronic supplementary information
for
# Ionic conductivity of deep eutectic solvents: The role of orientational dynamics and glassy freezing

Daniel Reuter, Catharina Binder, Peter Lunkenheimer* and Alois Loidl

*Experimental Physics V, Center for Electronic Correlations and Magnetism, University of Augsburg, 86135 Augsburg, Germany.*

*\* Corresponding author. E-mail: peter.lunkenheimer@physik.uni-augsburg.de*

## DSC results

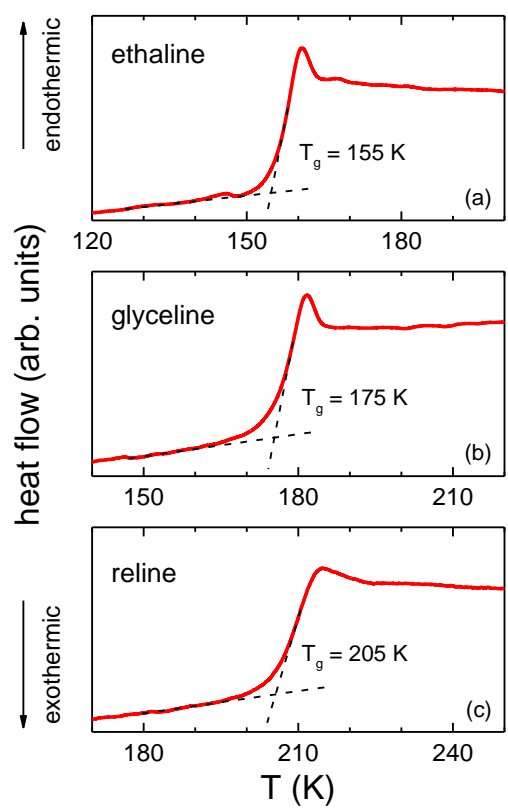

Fig. S1 DSC results on the three DESs treated in the present work as obtained under heating with 10 K/min.

Figure S1 shows the temperature-dependent DSC heat flow for the three investigated DESs. The glass temperatures were estimated by the onset method as indicated by the dashed lines.

1